\documentstyle[epsf,prd,aps,floats]{revtex}

\newcommand {\be}{\begin{equation}}
\newcommand {\ee}{\end{equation}}
\newcommand {\bea}{\begin{eqnarray}}
\newcommand {\eea}{\end{eqnarray}}

\begin{document}
\twocolumn[\hsize\textwidth\columnwidth\hsize\csname@twocolumnfalse\endcsname

\title{Is it $e$ or is it $c$? Experimental Tests of Varying Alpha}

\author{Jo\~ao Magueijo$^a$, John D.
Barrow$^b$, and H\aa vard Bunes Sandvik$^a$}
\address{{\it a} Theoretical Physics, The Blackett Laboratory,
 Imperial College, Prince Consort Rd., London, SW7 2BZ, U.K. \\
{\it b} DAMTP, Centre for Mathematical Sciences, Cambridge
University, Wilberforce Rd., Cambridge CB3 0WA, U.K.}

\maketitle

\begin{abstract}
Is the recent evidence for a time-varying fine structure 'constant' 
$\alpha$ to be interpreted as a
varying $e$, $c$, $\hbar$, or a combination thereof?   We consider the simplest varying
electric charge and varying speed of light theories (VSL) and prove
that for the same type of dark matter they predict opposite senses of 
variation in $\alpha$ over cosmological times. We also show
that unlike varying $e$ theories,  VSL theories do not predict violations of the
weak equivalence principle (WEP). Varying $e$ theories which explain astronomical 
inferences of varying $\alpha$ predict WEP violations 
only an order of magnitude smaller than existing  E\"otv\"os experiment limits but 
could be decisively tested by  STEP. We finally exhibit a set of
atomic-clock and related experiments for which {\it all}
(hyperbolic) varying $\alpha$ theories predict non-null results.
They provide independent tests of the recent astronomical
evidence.
\end{abstract}

\pacs{PACS Numbers: *** }

] \renewcommand{\thefootnote}{\arabic{footnote}} \setcounter{footnote}{0}


The possibility that the fine structure constant $\alpha \equiv e^2/\hbar c$
might be a dynamical variable has attracted considerable attention following
the observations of Webb et al \cite{murphy,webb,webb2}. These observations
use a powerful new many-multiplet technique to extract a larger fraction of
the information encoded in quasar (QSO) absorption spectra at medium
redshift than traditional doublet studies. They study energy differences
between relativistic transitions to many different ground states and compare
observations with laboratory measurements and many-body computations. The
continuing trend of these results is that the value of $\alpha $ was 
{\it lower} in the past, with $\Delta \alpha /\alpha =-0.72\pm 0.18\times
10^{-5}$ for $z\approx 0.5-3.5.$ 

It is clear that such a discovery, if
correct, has important implications for the foundations of physics, but
pinpointing the precise conclusions to be drawn is more controversial. While
it is reassuring that $\alpha $ was lower rather than higher in the past, so
that QED remains perturbative, the full implications for unification are
still unclear \cite{banks,guts}, and we shall not discuss them further here.
These results also raise the question: which of $e$, $\hbar $ and $c$ might
be responsible for any observed change in $\alpha $ and what operational
meaning should be attributed to such a determination? Undoubtedly, in the
sense of \cite{dov}, one has to make an operationally ``meaningless'' choice
of which dimensional constant is to become a dynamical variable. Yet, in
practice this choice is never arbitrary; it is clearly dictated by
simplicity once the detailed dynamics of the theory have been established.
Here, we argue that the dynamics have unambiguous observational
implications: a combination of experiment and simplicity therefore
selects one member
of a dimensionless combination ($\alpha $) of dimensional constants ($e$, $%
\hbar $ and $c$) to which we should preferentially ascribe its space-time
variation. We will present a number of clear experimental tests which can
distinguish rival theories of $\alpha $ variation which are expressed
through explicit change in $e$ or $c$. Existing theories will be used as
examples.

Several theoretical contexts for the Webb et al results have been explored.
Sandvik, Magueijo and Barrow \cite{bsbm} have recently proposed a varying
electric charge theory (BSBM), inspired by an earlier construction of
Bekenstein \cite{bek2}. A supersymmetric version of this theory was created
in \cite{olive}. Various dilatonic alternatives, in which all coupling
constants vary as a function of a single field, may also be considered
(including dilaton couplings to the cosmological constant \cite{olive}).
Other candidates to explain variations in $\alpha $ are the varying speed of
light (VSL) theories~\cite{moffat93,am,ba,bm,barmag98,covvsl}, which also
offer an alternative to inflation for solving cosmological problems. As
archetypal examples we take the BSBM theory \cite{bsbm} and the VSL theory
presented in \cite{covvsl}. By introducing an appropriate change of units we
can turn VSL into a constant $c$ theory, but the dynamics will then look
unnecessarily complicated; likewise BSBM can be rephrased as a constant $e$,
varying $c$ theory, with a concomitant increase in complexity. This is why
we say that BSBM is a varying $e$ theory while the theory in \cite{covvsl}
is a VSL theory: dynamics fixes the choice. Crucially, the dynamics also
have unambiguous observational implications. We will show that with standard
dark matter VSL predicts an {\it increasing} $\alpha $, as a function of
cosmological time. By contrast, BSBM predicts a {\it decreasing} $\alpha $,
a conclusion which can only be reversed by a different choice of dark matter
composition, as explained in \cite{bsbm}. This is a striking difference, but
pending the determination of the nature of the dark matter one can use both
BSBM and VSL to fit the Webb et al results. The same remark applies to other
cosmological tests, such as constraints arising from the cosmic microwave
background (CMB) and Big Bang Nucleosynthesis (BBN) \cite{avelino,bat}.

However, 
BSBM and VSL theories also make different predictions regarding {\it spatial}
variations in $\alpha $ near massive objects. Due to these variations all
changing-$\alpha $ theories predict a 'fifth force' effect \cite
{bsbm,bek2,olive,zal,moffatal}, but we will see that the exact details can
distinguish between BSBM and VSL. In BSBM theory the fifth force induces an
anomalous acceleration which, unlike gravity, depends on the material
composition of the test particle and so violates the weak equivalence
principle (WEP). The VSL theories, on the other hand, are consistent with
the WEP, as first noted by Moffat\cite{moffatal}.

The exact level of WEP violation predicted by BSBM depends upon an unsolved
problem in nuclear and hadronic physics: how much of the mass-energy of
nuclei is of electrostatic nature? As yet, there is no reliable answer to
this question \cite{pdg} but we can still estimate the magnitude of WEP
violation, which reveals that the BSBM theory is marginally consistent with
current E\"{o}tv\"{o}s experiments. However, the next generation of WEP
tests, such as the STEP project \cite{step}, 
will easily be sensitive enough to detect
violations of the WEP as predicted by BSBM even by the most conservative
estimates. Should violations be observed, it should be seen as a success for
varying $e$ theories. If not, then we must narrow our interest to VSL
theories in order to accommodate observational signals of varying $\alpha $.
Thus, space experiments such as STEP 
 can provide an independent experimental
test of any astronomical  evidence for varying $\alpha$, and decide
between a varying $e$ or $c$ interpretation.

We start by describing briefly the two theories to be used as exemplars. In
the BSBM varying $\alpha $ theory, the quantities $c$ and $\hbar $ are taken
to be constant, while $e$ varies as a function of a real scalar field $\psi
, $ with $e=e_0e^\psi $. As shown in \cite{string}, it is possible to
rewrite this theory in such a way that $\psi $ only couples to the free
electromagnetic lagrangian ${\cal L}_{em}$. The field tensor $f_{\mu \nu
}=\partial _\mu a_\nu -\partial _\nu a_\mu $ and the covariant derivatives $%
D_\mu =\partial _\mu +ie_0a_\mu $ then do not contain $\psi $, and the
action takes the form: 
\begin{equation}
S=\int d^4x\sqrt{-g}\left( {\cal L}_g+{\cal L}_{mat}+{\cal L}_\psi +{\cal L}%
_{em}e^{-2\psi }\right) ,
\end{equation}
where ${\cal L}_\psi =-{\frac \omega 2}\partial _\mu \psi \partial ^\mu \psi 
$, ${\cal L}_{em}=-\frac 14f_{\mu \nu }f^{\mu \nu }$, and ${\cal L}_{mat}$
(the lagrangian of all matter fields apart from ${\cal L}_{em}$) does not
depend on $\psi $. The gravitational lagrangian is the usual ${\cal L}_g=%
\frac 1{16\pi G}R$, with $R$ the curvature scalar.

In contrast, the covariant VSL theory proposed in \cite{covvsl} assumes that 
$c$ varies, and builds the simplest dynamics on this premise, which is
equivalent to a choice of a system of units. It assumes that $c=c_0e^\chi $
(with $\chi $ another real scalar field) and that the {\it full} matter
Lagrangian ${\cal L}_M$ does not contain $\chi $. Up to a free parameter, $q$%
, this assumption fixes how all matter couplings scale with $c$; in
particular, one has for all interactions $i$ associated with gauge charges $%
e_i$ that $\alpha _i\propto e_i\propto \hbar c\propto c^q$. The action is 
\cite{values} 
\begin{equation}
S=\int d^4x\sqrt{-g}\left( {\cal L}_g+{\cal L}_\chi +{\cal L}_Me^{b\chi
}\right) ,
\end{equation}
with ${\cal L}_\chi =-{\frac \omega 2}\partial _\mu \chi \partial ^\mu \chi $%
, and ${\cal L}_g$ is as given above. It was shown in \cite{covvsl} that
only when $b+q\neq 0$ can these theories be conformally mapped into dilaton
theories, and into Brans-Dicke theories only when $q=0$ . This theory has an
obvious novelty when compared to BSBM: all $\alpha _i$ are variable.
However, recent cosmological variations in non-electromagnetic $\alpha _i$
are beyond the reach of current direct astrophysical observations. Hence for
the purpose of this Letter we shall ignore their consequences.

Varying the action with respect to the metric leads to straightforward
generalizations of Einstein's equations \cite{covvsl,bsbm}. Variation with
respect to the new scalar fields leads to dynamical equations for $\alpha $.
For small variations, $\delta \alpha /\alpha \ll 1,$ these are: 
\begin{equation}
\Box {\frac{\delta \alpha }\alpha }=\frac 4\omega {\cal L}_{em}
\label{boxpsi}
\end{equation}
for BSBM, and 
\begin{equation}
\Box {\frac{\delta \alpha }\alpha }=-\frac{bq}\omega {\cal L}_M
\label{boxpsi1}
\end{equation}
for VSL. In both cases the right-hand side is zero for relativistic matter,
predicting negligible variations in $\alpha $ during the radiation-dominated
cosmological epoch. Two striking differences appear in the matter epoch,
when the RHS becomes non-negligible, in both the coupling parameters and the
driving source ${\cal L}$. The requirement that the fields $\chi $ and $\psi 
$ have a positive definite energy forces $\omega >0$. This fixes the sign of
the coupling for BSBM ($4/\omega $) but not for VSL ($-bq/\omega $). The
source ${\cal L}$ is also different for each theory and is parameterized by
different ratios determined by the dark matter: $\zeta ={\cal L}_{em}/\rho $
for BSBM, and $\xi ={\cal L}_M/\rho $ for VSL.

The value of $\zeta $ for baryonic and dark matter has been disputed \cite
{zal,olive,bsbm}. It is the difference between the percentage of mass in
electrostatic and magnetostatic forms. As explained in \cite{bsbm}, we can
at most {\it estimate} this quantity for neutrons and protons, with $\zeta
_n\approx \zeta _p\sim 10^{-4}$. We may expect that for baryonic matter $%
\zeta \sim 10^{-4}$, with composition-dependent variations of the same
order. The value of $\zeta $ for the dark matter, for all we know, could be
anything between -1 and 1. Superconducting cosmic strings, or magnetic
monopoles, display a {\it negative} $\zeta $, unlike more conventional dark
matter. On the other hand it was argued in \cite{covvsl} that the value of $%
\xi $ (characterizing the VSL dynamics in the matter epoch) is $-1$ for all
non-relativistic matter. This is equivalent to requiring that
non-relativistic matter is dominated by its potential energy (including rest
mass)$\ $rather than by its kinetic energy $T$. We shall use this fact in
the rest of the paper although it is not essential for most of what follows.

It is clear that the only way to obtain a cosmologically increasing $\alpha $
in BSBM is with $\zeta <0$, i.e with unusual dark matter, in which magnetic
energy dominates over electrostatic energy. In \cite{bsbm} we showed that
fitting the Webb et al results requires $\zeta _m/\omega =-2\pm 1\times
10^{-4}$, where $\zeta _m$ is weighted by the necessary fractions of dark
and baryonic matter. On the other hand VSL theory fits the Webb et al
results with $bq/\omega =-8\times 10^{-4}$, for all types of dark matter.
Hence, if we were to determine that $\zeta >0$ for the dark matter in the
universe, we could experimentally rule out BSBM but not VSL. This is just
one way in which the question in the title of this paper could be answered.

However, pending identifying the dark matter, we may still answer this
question by looking at spatial variations in $\alpha $. In all causal
varying-$\alpha $ theories defined by a wave equation the observed redshift
dependence of $\alpha $ requires there also to be spatial variations near
compact massive bodies \cite{bhvsl,bsbm}. The relevant equations may be
obtained by dropping the time dependence in (\ref{boxpsi}) and (\ref{boxpsi1}%
). Then, a linearized spherically symmetric solution in the vicinity of an
object with mass $M_s$ and $\zeta =\zeta _s$ is 
\begin{equation}
{\frac{\delta \alpha }\alpha }=-{\frac{\zeta _s}\omega }{\frac{M_s}{\pi r}}%
\approx 2\times 10^{-4}{\frac{\zeta _s}{\zeta _m}}{\frac{M_s}{\pi r}}
\label{alphar}
\end{equation}
for BSBM 
\begin{equation}
{\frac{\delta \alpha }\alpha }=-{\frac{bq}\omega }{\frac{M_s}{4\pi r}}%
\approx 2\times 10^{-4}{\frac{M_s}{\pi r}}\;.  \label{alphar1}
\end{equation}
for VSL. We note that the level of spatial variations in BSBM, given \cite
{webb,webb2}, depends on the nature of the dark matter (the ratio $\zeta
_s/\zeta _m$), whereas for VSL it does not. In VSL, $\alpha $ {\it increases}
near compact objects (with decreasing $c$ if $q<0$, with increasing $c$ if $%
q>0$) but in BSBM $\alpha $ {\it decreases} (since $\zeta _m<0$ and $\zeta
_s>0$). In VSL theories, near a black hole $\alpha $ could become much
larger than 1, so that electromagnetism would become non-perturbative with
dramatic consequences for the physics of black holes. In BSBM precisely the
opposite happens: electromagnetism switches off.

Spatial variations lead to a number of observable effects which sharply
distinguish between VSL and BSBM. Most obviously $\alpha $ could be measured
in absorption lines from compact objects, as explained in \cite{covvsl,bsbm}%
. More subtly, alpha gradients induce a 'fifth force' effect. In order to
compute this force one must model $\zeta $ or $\xi $ for test bodies. In
BSBM the test-particle lagrangian may be split as ${\cal L}_t={\cal L}%
_m+e^{-2\psi }{\cal L}_{em}$. Variation with respect to the metric leads to
a similar split of the stress-energy tensor, producing an energy density of
the form $\rho ((1-\zeta _t)+\zeta _te^{-2\psi }),$ and so a mass of $%
m((1-\zeta _t)+\zeta _te^{-2\psi }),$ (assuming electric fields dominate).
In order to preserve their ratios of $\zeta _t={\cal L}_{em}/\rho $ test
particles may thus be represented by 
\begin{equation}
{\cal L}(y)=-\int d\tau \;m((1-\zeta _t)-\zeta _te^{-2\psi })[-g_{\mu \nu }%
\dot x^\mu \dot x^\nu ]^{\frac 12}{\frac{\delta (x-y)}{{\sqrt{-g}}}}
\end{equation}
where over-dots are derivatives with respect to the proper time $\tau $.
This leads to equations of motion: 
\begin{equation}
\ddot x^\mu +\Gamma _{\alpha \beta }^\mu \dot x^\alpha \dot x^\beta +{\frac{%
2\zeta _te^{-2\psi }}{(1-\zeta _t)-\zeta _te^{-2\psi }}}\partial ^\mu \psi =0
\end{equation}
which in the non-relativistic limit (with $\zeta _t\ll 1$) reduce to 
\begin{equation}
{\frac{d^2x^i}{dt^2}}=-\nabla _i\phi -2\zeta _t\nabla _i\psi \;,
\end{equation}
where $\phi $ is the gravitational potential. Thus we predict an anomalous
acceleration: 
\begin{equation}
a={\frac{M_s}{r^2}}{\left( 1+{\frac{\zeta _s\zeta _t}{\omega \pi }}\right) }
\label{acc}
\end{equation}
Violations of the WEP occur because $\zeta _t$ is substance dependent. For
two test bodies with $\zeta _1$ and $\zeta _2$ the E\"otv\"os parameter is: 
\begin{equation}
\eta \equiv {\frac{2|a_1-a_2|}{a_1+a_2}}={\frac{\zeta _s|\zeta _1-\zeta _2|}{%
\omega \pi }.}
\end{equation}
This can be written more conveniently as the product of the following 3
factors: 
\begin{equation}
\eta =\left( {\frac{\zeta _E|\zeta _1-\zeta _2|}{\pi \zeta _p}}\ \right)
\left( {\frac{\zeta _p}{\zeta _m}}\right) \left( {\frac{\zeta _m}\omega }%
\right) .
\end{equation}
The last factor is the coupling that determines cosmological time variations
in $\alpha $, and using the results\cite{webb,webb2} is best fitted to be $%
\zeta _m/\omega \approx -10^{-4}$. If we take $\zeta _n\approx \zeta
_p\approx |\zeta _p-\zeta _n|={\cal O}(10^{-4})$ then for typical substances
the first factor is $\approx 10^{-5}$. Hence, we need $\zeta _m={\cal O}(1)$
to produce $\eta ={\cal O}(10^{-13}),$just an order of magnitude below
existing experimental bounds.

In contrast to this VSL theories predict that for all test particles 
\begin{equation}
{\cal L}(y)=-\int d\tau \;me^{b\chi }[-g_{\mu \nu }\dot x^\mu \dot x^\nu ]^{%
\frac 12}{\frac{\delta (x-y)}{{\sqrt{-g}}},}
\end{equation}
where we have assumed $\xi =-1$. This leads to an anomalous acceleration of
equal magnitude for all test particles, so that there are no WEP violations.
This new acceleration does imply corrections to the standard tests of
general relativity, such as the precession of Mercury's perihelion, light
deflection and radar echo time-delay \cite{will,bhvsl}. These were studied
in \cite{bhvsl} and impose the undemanding constraint of $b^2/\omega
<10^{-2} $ \cite{note2}. Therefore we conclude that an increase of about an
order of magnitude in the experimental sensitivity to non-zero $\eta $ would
decide between the BSBM and VSL theories.

Webb et al \cite{murphy,webb,webb2} caution that their results might be due
to some uninvestigated systematic effect. For this reason it is important to
seek independent observational verification. Direct measurement of WEP
violations at the predicted level could be seen as a direct confirmation of
the source of the astronomical results. Spatial fluctuations in $\alpha $
could also be directly mapped from spectroscopy of lines formed in very
compact objects or their accretion disks \cite{bsbm,bhvsl}. But more
realistically we note that Earth-based atomic-clock experiments could also
measure these fluctuations. Atomic clocks tick at a rate $\tau ^{-1}\propto
(E_e\alpha ^2)/\hbar $, where $E_e$ is the electron rest energy. Hence
atomic-clock experiments able to measure gravitational redshifts will suffer
from an extra effect: in BSBM theories these clocks tick slower in
gravitational wells, with $\tau \propto 1/\alpha ^2$, whereas in VSL $\tau
\propto 1/c^{2q+1}\propto 1/\alpha ^{2+1/q}$. Any hyperbolic varying-$\alpha 
$ theory explaining \cite{webb,webb2} should predict a similar effect.

In general, any gravitational-redshift experiment should be sensitive to a
varying $\alpha $. For example, the Pound-Rebka-Snyder experiment uses the
M\"{o}ssbauer effect to produce a narrow resonance line from $\gamma $-ray
photons emitted by radioactive isotopes. The effect has been used to observe
gravitational redshifts, but the emitted photon's energy also depends upon $%
\alpha $. For small variations in $\alpha $ the energy shift is $\delta
E/E=C\delta \alpha /\alpha $ with $C$ of order 1 (but not very well known).
A similar effect will occur in experiments using Rydberg lines, with a shift
in wavelength $\delta \lambda /\lambda =-2(\delta \alpha /\alpha )$ (for
both VSL and BSBM theories). Once the photon is emitted, varying-$\alpha $
theories predict no extra redshift for free-flying photons (since ${\cal L}=0
$ for photons). However, the observed gravitational redshift of frequencies
takes the form $\delta \nu /\nu =(1+\alpha _{PPN})\delta \phi $, with a
non-zero PPN parameter $\alpha _{PPN}$ induced {\it at emission}. Using (\ref
{alphar}) we find that for BSBM theory $\alpha _{PPN}=2\zeta _s/(\pi \omega )
$, with the quasar data \cite{webb,webb2} then implying that $\alpha
_{PPN}\approx 10^{-8}$. For VSL theory care must be taken, because $\delta
\lambda /\lambda $, $\delta \nu /\nu $ and $\delta E/E$ are distinct
quantities. Defining $\alpha _{PPN}$ in terms of frequency in the
conventional way and using (\ref{alphar1}) we have that $\alpha
_{PPN}\approx (2+q^{-1})bq/(4\pi \omega )\approx -(2+q^{-1})10^{-5}$. Hence
BSBM theory predicts a stronger redshift than general relativity, with
corrections of order $10^{-8}$. If $q\ll 1,$ VSL theory predicts a weaker
redshift effect with corrections of order $10^{-5}$; but this conclusion is
changed if $q\approx -1/2$. Both BSBM and VSL theories are consistent with
the current bound of $\alpha _{PPN}<10^{-4}$ \cite{will}. Any causal varying-%
$\alpha $ theory should predict a non-zero correction to the relativistic
redshift formulae.

In summary, we have explained how a combination of experiment and common
sense may distinguish a varying $c$ from a varying $e$. Using only minimal
versions of such theories, we have shown how they can be distinguished by
weak equivalence principle violations, by the type of dark matter required
to give the variations inferred from quasar observations\cite{webb,webb2},
and by gravitational-redshift experiments. In non-minimal varying-$e$ and -$c
$ theories, the distinguishing observational signatures should be even more
obvious. For instance, if Lorentz invariance were found to be broken, \cite
{amel,stephon}, then a varying-$c$ theory would provide a better framework
for expressing variations in $\alpha $. 

Finally, we note that the
experiments proposed in this paper are by no means the only discriminators
between varying-$e$ and -$c$ expressions of a varying $\alpha $. In \cite
{davis} the authors examined black hole thermodynamics, by changing the
values of $e$ and $c$ in their description of black hole thermodynamics
(which, however, may be too simplistic \cite{bhvsl}). In this context they
found that interpreting a varying $\alpha $ as varying $e$ or $c$ leads to
opposite black-hole dynamics, with a varying-$e$ contradicting the second
law of thermodynamics. In principle one could test whether or not black hole
areas always increase with time in the next generation of gravitational-wave
observatories. Like the various experiments described in this paper, this
is experimentally unambiguous, since the ratio of two areas is dimensionless.

{\bf Acknowledgements} HBS thanks the Research Council of Norway for
financial support. We thank S. Alexander, P. Davies, T. Davis, G. Dvali, J.
Moffat, D. Mota, M. Murphy, K. Olive, M. Pospelov, G. Shore, J. Webb and M.
Zaldarriaga for discussions.


\begin{references}
\bibitem{values}  In the present version of \cite{covvsl} we have set $a=0$
and $\omega =\kappa /(8\pi )$.
\bibitem{note2}  In \cite{bhvsl} we actually used $\xi =-1/2$, but it is
easy to adapt these results to $\xi =-1$.
\end{references}
\end{document}